\documentclass[12pt]{article}
\usepackage{amssymb}
\usepackage{graphicx}
\usepackage[cp1251]{inputenc}
\usepackage{rotating}

\begin{document}
 \noindent {\footnotesize\it Astronomy Reports, 2024, Vol. 68, No. 9, pp. 878--885}
 \newcommand{\dif}{\textrm{d}}

 \noindent
 \begin{tabular}{llllllllllllllllllllllllllllllllllllllllllllll}
 & & & & & & & & & & & & & & & & & & & & & & & & & & & & & & & & & & & & & &\\\hline\hline
 \end{tabular}

  \vskip 0.5cm
  \bigskip
 \bigskip
\centerline{\large\bf Kinematics of Fainst Stars of the Sco-Cen Association}
\centerline{\large\bf According to Gaia Catalog}
 \bigskip
 \bigskip
  \centerline { 
   V. V. Bobylev\footnote [1]{bob-v-vzz@rambler.ru},  A. T. Bajkova}
 \bigskip
 \centerline{\small\it Pulkovo Astronomical Observatory, Russian Academy of Sciences, St. Petersburg, 196140 Russia}
 \bigskip
 \bigskip
{{\bf Abstract}--The kinematic properties of the Sco-Cen association have been studied using the spatial velocities of young stars. New kinematic age estimates for the three components of the association with the age of UCL and LCC being $17.7\pm2.4$~Myrs and the age of US being $6.4\pm1.7$~Myrs have been obtained. The parameters of the residual velocities US, UCL, and LCC ellipsoid have been estimated.
 }

\bigskip
\section{INTRODUCTION}
The OB2 stellar association Sco-Cen stretched in longitude almost along the entire fourth galactic quadrant is located near the Sun [1, 2]. It consists of three main structural units: US (Upper Scorpius), UCL (Upper Centaurus-Lupus), and LCC (Lower Centaurus-Crux) with average distances of 145, 140, and 118 pc, respectively [3, 4]. The age of UCL and LCC is
estimated at 16--20 million years while the age of
younger US is less than 10 million years. These age
estimates are obtained by fitting the positions of stars on the Hertzsprung-Russell diagram to suitable theoretical isochrones [5--7].

The Sco-Cen association contains several thousand stars. Its members are both massive stars of spectral classes B2 and later (hence, it is sometimes designated as Sco OB2 [8]), as well as low-mass stars of spectral classes K--M (T Tauri-type stars). With a
complex, patchy structure, the Sco-Cen association
includes dozens of smaller associations of low-mass
stars, open star clusters, and small stellar groupings of
various ages [9, 10]. For example, $\rho$~Oph/L1688,
Sco, Sco, and others in US; V1062 Sco, Lupus, and others in UCL; and $\epsilon$~Cham, $\eta$~Cham, and others in LCC. Therefore, at present, some authors prefer to indicate an interval of 1--20 million years when considering the age of the Sco-Cen association.

A study of young low-mass stars in the Sco-Cen association showed [11] that the birth of such stars in US is in good agreement with the sequential star formation model. According to this model, the explosions of massive stars as supernovae in one part of the
association trigger the compression of hydrogen
clouds and the onset of star formation in neighboring
regions. Moreover, it is believed that relatively recent explosions of the most massive stars in the Sco-Cen association led to the formation of the North Polar Snort [12], and they are also responsible for the occurrence of the Local Bubble [13].

An interesting and important kinematic property of
the Sco-Cen association is its expansion effect first
discovered in [2, 14]. The value of the angular
expansion coefficient of the association has been
repeatedly clarified by various authors [15--18]. Using
this coefficient, $K\sim$50 km/s/kpc, the kinematic age of
the association is estimated at $\sim$20 million years.
However, the authors of the work [19] found no evidence
of expansion of the Sco-Cen association.

The emergence of small stellar associations close to
the Sun, such as $\beta$~Pic ($\sim$20 million years [20, 21]) and
TW Hya ($\sim$10 million years [22, 23]), apparently
closely related to the evolution of the Sco-Cen OB
association. There is satisfactory agreement between
the age estimates obtained by different methods:
(a) isochronous, (b) based on the analysis of the lithium
content in stars, and (c) kinematic.

The spatial size of the Sco-Cen association is significant:
it is stretched along galactocentric direction
by $\sim$300 pc. Authors of [24] associate the formation
of this association with the impact of the galactic spiral
density wave on the parent gas-dust cloud. In the case
of using kinematic methods, it is desirable to take into
account the impact of the spiral density wave in the velocities of stars. This approach was implemented, for example, in [17, 25]. The only problem is the poor knowledge of the specific values of the parameters of the spiral wave in the Galaxy.

The spatial and kinematic properties of relatively bright stars in the Sco-Cen association have been studied in detail using data from the Hipparcos catalogue
[26], for example, in [3, 17--19, 24, 25]. The Gaia space project provides the opportunity to study fainter and more numerous stars [27, 28].

The aim of this paper is to study the kinematic properties of young faint stars in the Sco-Cen association using the values of their trigonometric parallaxes and proper motions from the Gaia EDR3 catalogue [29]. For this, we consider the stars of the Sco-Cen association from the list of [29].

\section{METHODS}
We use a rectangular coordinate system centered on the Sun, where the axis $x$ directed towards the galactic center, axis $y$~--- towards the galactic rotation and axis $z$~--- to the north pole of the Galaxy. Then $x=r\cos l\cos b,$ $y=r\sin l\cos b$ and $z=r\sin b,$ where $r=1/\pi$ is the heliocentric distance of the star in kpc, which we calculate through the trigonometric parallax of the star $\pi$ in mas (milliarcseconds).

The radial velocity is known from observations $V_r$ and two projections of tangential velocity $V_l=4.74r\mu_l\cos b$ and $V_b=4.74r\mu_b,$ directed along galactic longitude $l$ and latitude $b$ respectively, expressed in km/s. Here, the coefficient 4.74 is the ratio of the number of kilometers in an astronomical unit to the number of seconds in a tropical year. Components of proper motion $\mu_l\cos b$ and $\mu_b$ expressed in mas/year.

Through components $V_r, V_l, V_b,$ velocities are calculated $U,V,W,$ where $U$ is the velocity directed from the Sun to the center of the Galaxy, $V$ in the direction of rotation of the Galaxy and $W$~--- to the north galactic pole:
 \begin{equation}
 \begin{array}{lll}
 U=V_r\cos l\cos b-V_l\sin l-V_b\cos l\sin b,\\
 V=V_r\sin l\cos b+V_l\cos l-V_b\sin l\sin b,\\
 W=V_r\sin b                +V_b\cos b.
 \label{UVW}
 \end{array}
 \end{equation}
Thus, velocityes $U,V,W$ directed along the corresponding coordinate axes $x,y,z$.

\subsection{Construction of Orbits of Stars}
To construct the orbits of stars in a coordinate system rotating around the center of the Galaxy, we use the epicyclic approximation [13, 30]:
 \begin{equation}
 \renewcommand{\arraystretch}{1.8}
 \begin{array}{lll}\displaystyle
 x(t)= x_0+{U_0\over \displaystyle \kappa}\sin(\kappa t)+{\displaystyle V_0\over \displaystyle 2B}(1-\cos(\kappa t)),  \\
 y(t)= y_0+2A \biggl(x_0+{\displaystyle V_0\over\displaystyle 2B}\biggr) t
       -{\displaystyle \Omega_0\over \displaystyle B\kappa} V_0\sin(\kappa t)
       +{\displaystyle 2\Omega_0\over \displaystyle \kappa^2} U_0(1-\cos(\kappa t)),\\
 z(t)= {\displaystyle W_0\over \displaystyle \nu} \sin(\nu t)+z_0\cos(\nu t),
 \label{EQ-Epiciclic}
 \end{array}
 \end{equation}
where $t$~--- time in million years (we proceed from the ratio 1 pc/1 million years = 0.978 km/s), $A$ and $B$ are the Oort constants; $\kappa=\sqrt{-4\Omega_0 B}$ is epicyclic frequency; $\Omega_0$ is the angular velocity of the galactic rotation of the local rest standard, $\Omega_0=A-B$; and $\nu=\sqrt{4\pi G \rho_0}$ is the frequency of vertical vibrations, where $G$ is the gravitational constant, and $\rho_0$ is the stellar density in the solar vicinity.

Parameters $x_0,y_0,z_0$ and $U_0,V_0,W_0$ in the system of equations (2) denote the modern positions and velocities of stars, respectively. The rise of the Sun above
galactic plane $h_\odot$ taken equal to 16 pc according to [31]. Velocities $U,V,W$ calculate relative to the local standard of rest using the values
 \begin{equation}
(U_\odot,V_\odot,W_\odot)=(11.1,12.2,7.3~\hbox {km/s}
 \end{equation}
obtained in [32]. We accepted $\rho_0=0.1~M_\odot/$pc$^3$~[33], what gives $\nu=74$~km/s/kpc. We use the following values of the Oort constants: $A=16.9$~km/s/kpc and $B=-13.5$~km/s/kpc, close to modern estimates. An overview of such estimates can be found, for example, in [34].

The system of equations (2) makes it possible to calculate the position of a star at any specified moment in time. The values of all constants are defined, and the solution is found by substituting the moment of time.

\subsection{Residual Velocity Ellipsoid}
To estimate the dispersions of residual velocities of stars we use the well-known method [35], where six second-order moments are considered:
$a=\langle U^2\rangle-\langle U^2_\odot\rangle,$
 $b=\langle V^2\rangle-\langle V^2_\odot\rangle,$
 $c=\langle W^2\rangle-\langle W^2_\odot\rangle,$
 $f=\langle VW\rangle-\langle V_\odot W_\odot\rangle,$
 $e=\langle WU\rangle-\langle W_\odot U_\odot\rangle$ and
 $d=\langle UV\rangle-\langle U_\odot V_\odot\rangle,$
which are the coefficients of the surface equation
 \begin{equation}
 ax^2+by^2+cz^2+2fyz+2ezx+2dxy=1,
 \end{equation}
as well as components of the symmetric tensor of moments of residual velocities
 \begin{equation}
 \left(\matrix {
  a& d & e\cr
  d& b & f\cr
  e& f & c\cr }\right).
 \label{ff-5}
 \end{equation}
The following six equations are used to determine the values of this tensor:
\begin{equation}
 \begin{array}{lll}
 V^2_l= a\sin^2 l+b\cos^2 l\sin^2 l  -2d\sin l\cos l,
 \label{EQsigm-1}
 \end{array}
 \end{equation}
\begin{equation}
 \begin{array}{lll}
 V^2_b= a\sin^2 b\cos^2 l+b\sin^2 b\sin^2 l   +c\cos^2 b  \\
 -2f\cos b\sin b\sin l   -2e\cos b\sin b\cos l    +2d\sin l\cos l\sin^2 b,
 \label{EQsigm-2}
 \end{array}
 \end{equation}
\begin{equation}
 \begin{array}{lll}
 V_lV_b= a\sin l\cos l\sin b   +b\sin l\cos l\sin b\\
 +f\cos l\cos b-e\sin l\cos b   +d(\sin^2 l\sin b-\cos^2\sin b),
 \label{EQsigm-3}
 \end{array}
 \end{equation}
\begin{equation}
 \begin{array}{lll}
 V^2_r= a\cos^2 b\cos^2 l+b\cos^2 b\sin^2 l  +c\sin^2 b \\
 +2f\cos b\sin b\sin l   +2e\cos b\sin b\cos l +2d\sin l\cos l\cos^2 b,
 \label{EQsigm-4}
 \end{array}
 \end{equation}
\begin{equation}
 \begin{array}{lll}
 V_b V_r=-a\cos^2 l\cos b\sin b   -b\sin^2 l\sin b\cos b+c\sin b\cos b\\
 +f(\cos^2 b\sin l-\sin l\sin^2 b)   +e(\cos^2 b\cos l-\cos l\sin^2 b) \\
 -d(\cos l\sin l\sin b\cos b  +\sin l\cos l\cos b\sin b),
 \label{EQsigm-5}
 \end{array}
 \end{equation}
\begin{equation}
 \begin{array}{lll}
 V_l V_r=-a\cos b\cos l\sin l + b\cos b\cos l\sin l \\
    +f\sin b\cos l-e\sin b\sin l  + d(\cos b\cos^2 l-\cos b\sin^2 l),
 \label{EQsigm-6}
 \end{array}
 \end{equation}
which are solved by the least-squares method for six unknowns $a, b, c, f, e, d.$ Then the eigenvalues of the tensor (5) $\lambda_{1,2,3}$ are found from the solution of the secular equation
\begin{equation}
 \left|\matrix
 {
a-\lambda&          d&        e\cr
       d & b-\lambda &        f\cr
       e &          f&c-\lambda\cr
 }
 \right|=0.
 \label{ff-7}
 \end{equation}
The eigenvalues of this equation are equal to the reciprocal values of the squared semi-axes of the of velocity moment ellipsoid and, at the same time, the squared semi-axes of the residual velocity ellipsoid:
 \begin{equation}
 \begin{array}{lll}
 \lambda_1=\sigma^2_1, \lambda_2=\sigma^2_2, \lambda_3=\sigma^2_3,\qquad
 \lambda_1>\lambda_2>\lambda_3.
 \end{array}
 \end{equation}
Directions of the main axes of the tensor (12)  $L_{1,2,3}$ and $B_{1,2,3}$ are found from the relations:
 \begin{equation}
 \tan L_{1,2,3}={{ef-(c-\lambda)d}\over {(b-\lambda)(c-\lambda)-f^2}},
  \end{equation}
 \begin{equation}
 \tan B_{1,2,3}={{(b-\lambda)e-df}\over{f^2-(b-\lambda)(c-\lambda)}}\cos L_{1,2,3}.
  \end{equation}
Errors of determination of $L_{1,2,3}$ and $B_{1,2,3}$ are estimated
according to the following scheme:
 \begin{equation}
 \renewcommand{\arraystretch}{2.2}
  \begin{array}{lll}
  \displaystyle
 \varepsilon (L_2)= \varepsilon (L_3)= {{\varepsilon (\overline {UV})}\over{a-b}},\\
  \displaystyle
 \varepsilon (B_2)= \varepsilon (\varphi)={{\varepsilon (\overline {UW})}\over{a-c}},\\
  \displaystyle
 \varepsilon (B_3)= \varepsilon (\psi)= {{\varepsilon (\overline {VW})}\over{b-c}},\\
  \displaystyle
 \varepsilon^2 (L_1)={\varphi^2 \varepsilon^2 (\psi)+\psi^2 \varepsilon^2 (\varphi)\over{(\varphi^2+\psi^2)^2}},\\
  \displaystyle
 \varepsilon^2 (B_1)= {\sin^2 L_1 \varepsilon^2 (\psi)+\cos^2 L_1 \varepsilon^2 (L_1)\over{(\sin^2 L_1+\psi^2)^2}},
 \label{ff-65}
  \end{array}
 \end{equation}
where
 $\varphi=\cot B_1 \cos L_1$ and $\psi=\cot B_1 \sin L_1.$
In this case, it is necessary to calculate three quantities in
advance  $\overline {U^2V^2}$, $\overline {U^2W^2}$ and $\overline {V^2W^2},$ then
 \begin{equation}
 \renewcommand{\arraystretch}{1.6}
  \begin{array}{lll}
  \displaystyle
 \varepsilon^2 (\overline {UV})= (\overline{U^2V^2}-d^2)/n, \\
  \displaystyle
 \varepsilon^2 (\overline {UW})= (\overline {U^2W^2}-e^2)/n, \\
  \displaystyle
 \varepsilon^2 (\overline {VW})= (\overline {V^2W^2}-f^2)/n,
 \label{ff-73}
  \end{array}
 \end{equation}
where $n$ is the number of stars. The errors of each axis
are estimated independently, except for $L_2$ and $L_3,$ the
errors of which are calculated using the same formula.

  \begin{table}[t]
  \caption[]{\small
  Parameters of the US, UCL and LCC centers obtained by us for the selected groups of stars $N$
 }
  \begin{center}  \label{T-1}    \small
  \begin{tabular}{|r|r|r|r|r|r|r|c|r|r|r|}\hline
   &&&&&&&  &&& \\
 Object & ${\overline x}_0$ & ${\overline y}_0$ & ${\overline z}_0$ &
          ${\overline U}$ & ${\overline V}$ & ${\overline W}$ &
            $N_{x,y,z}/N_{U,V,W}$ & ${\overline U_0}$ & ${\overline V_0}$ & ${\overline W_0}$  \\
    & pc & pc & pc & km/s & km/s & km/s &  &  km/s & km/s & km/s \\\hline

  US & 135.6 & $-20.8$ & 41.5 & $-4.9$ & $-16.2$ & $-6.9$ & 1250/377 & 6.2 & $-4.0$ & 0.4 \\
 UCL & 121.5 & $-48.2$ & 33.5 & $-5.7$ & $-19.1$ & $-4.9$ & 2831/379 & 5.4 & $-6.9$ & 2.4 \\
 LCC &  75.7 & $-93.0$ & 22.9 & $-7.3$ & $-20.9$ & $-6.1$ & 3451/375 & 3.8 & $-8.7$ & 1.2 \\
 \hline
 \end{tabular}\end{center}
 {\small The initial values of the coordinates of the objects ${\overline x}_0,{\overline y}_0,{\overline z}_0$, average values of observed velocities relative to the Sun ${\overline U},{\overline V},{\overline W}$, and their
velocities relative to the local standard of rest
${\overline U_0},{\overline V_0},{\overline W_0}$.
$N_{x,y,z}$ is the number of stars used to calculate the average coordinates
and average velocities $N_{U,V,W}$.}
 \end{table}
  \begin{table}[t]
  \caption[]{\small
  Parameters of US, UCL and LCC centers obtained by other authors

 }
  \begin{center}  \label{T-111}    \small
  \begin{tabular}{|r|r|r|r|r|r|r|c|c|c|}\hline
    &&&&&&&& \\
Object & ${\overline x}_0$ & ${\overline y}_0$ & ${\overline z}_0$ &
           ${\overline U}$ & ${\overline V}$ & ${\overline W}$ &
          $N_\star$ &
             Ref \\
    & pc & pc & pc & km/s & km/s & km/s &  & \\\hline

  US & 138 & $-22$ & 49 & $-0.9$ & $-16.9$ & $-5.2$ & 120 & [36]\\
     &     &       &    & $-6.7$ & $-16.0$ & $-8.0$ & 155 & [24]\\
     &     &       &    & $-4.6$ & $-16.1$ & $-7.0$ & 469 & [29]\\

 UCL & 121 & $-69$ & 32 & $-7.9$ & $-19.0$ & $-5.7$ & 218 & [36]\\
     &     &       &    & $-6.8$ & $-19.3$ & $-5.7$ & 262 & [24]\\

 LCC & 61 & $-100$ & 14 & $-11.8$ & $-15.0$ & $-6.7$ & 179 & [36]\\
     &     &       &    & $ -8.2$ & $-18.6$ & $-6.4$ & 192 & [24]\\

UCL/LCC&   &       &    & $ -6.4$ & $-19.9$ & $-5.5$ & 542 & [29]\\
 \hline
 \end{tabular}\end{center}
 {\small
 The initial values of coordinates ${\overline x}_0,{\overline y}_0,{\overline z}_0$ and the average values of the observed velocities relative to the Sun ${\overline U},{\overline V},{\overline W}$ are given.
}
 \end{table}

\section{DATA}
The basis of our sample is made up of stars from the Sco-Cen association from the list [29]. The entire list contains over 10000 young stars with kinematic data
from the Gaia EDR3 catalog [28]. For some stars,
radial velocity measurements are available from various
sources. According to [29], $\sim$90\% of this list are stars of spectral classes M6--7 and later. Besides US, the combined UCL/LCC grouping, stars belonging to smaller clusters and associations are marked with special flags: V1062 Sco, Ophiuchus, and Lupus.

In this paper, we are interested in three main components of the Sco-Cen association: US, UCL, and LCC. From the list [29], we selected 1250 US stars with trigonometric parallaxes and proper motions, of which radial velocities are known for 377 stars. Stars
belonging to US were selected using the following two
flags: ``u'' in position 373 and `` '' in position 374 of
table1.dat from [29]. Stars belonging to UCL/LCC
were selected using the following two flags: ``c'' in
position 373 and `` '' in position 374. We divided the
UCL/LCC group into two parts with a border $l=325^\circ.$ The UCL contains 2831 stars, including 379 stars with known radial velocities. The LCC contains
3451 stars, of which 375 stars have known radial velocities. Stars with relative errors in determining trigonometric parallax of no more than 10\% were used.

Spatial velocities $U,V,$ and $W$ using the relations (1) can be calculated having all three measured velocities $V_l,V_b,$ and $V_r$. About 370 stars in each of the
selected US, UCL, or LCC groups can serve for these
purposes, as well as for constructing epicyclic orbits.
However, the situation with residual velocities $\sigma_1,\sigma_2$ and $\sigma_3$  (13) is different. At small distances from the Sun, as in our case, the errors in velocities and
(due to the high accuracy of measurements of parallaxes
and proper motions of stars from the Gaia satellite) are known to have on average a significantly lower level compared to the errors in $V_r$. Therefore, the results of both the joint solution of equations (6)--(11) and solutions using only the first three equations (6)--(8), i.e., without using radial velocities.

 \begin{figure} {\begin{center}
 \includegraphics[width=0.99\textwidth]{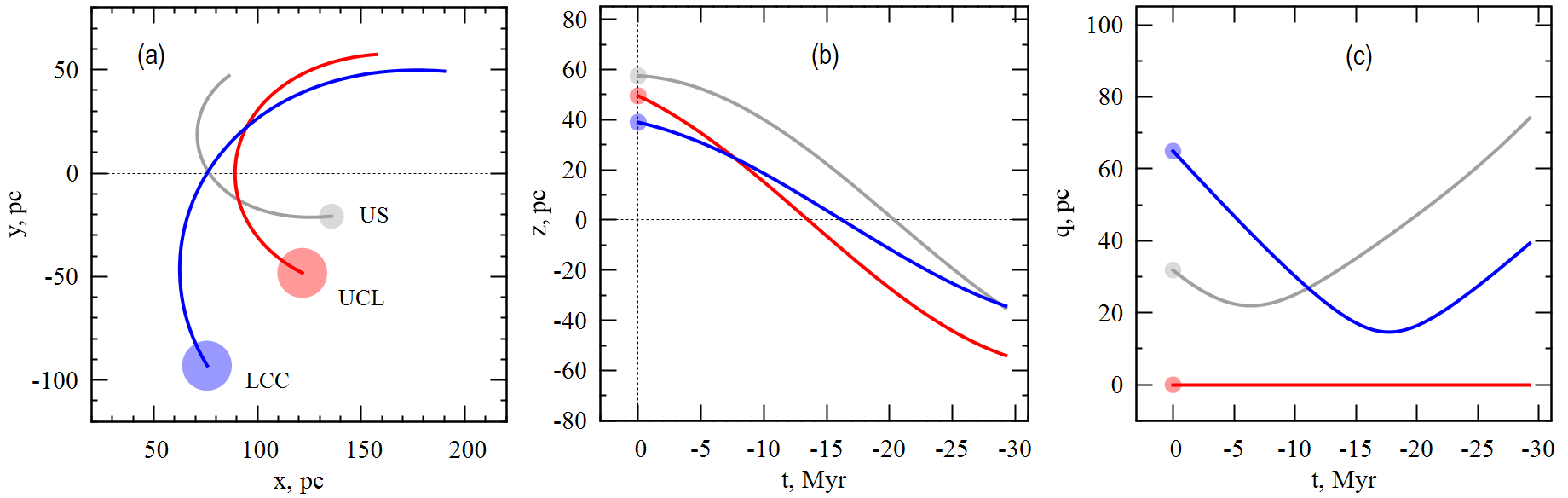}
 \caption{
(a) Centers of the three main components of the Sco-Cen association marked by circles of arbitrary size, projected ontomthe $xy$ galactic plane and their trajectories in the past on time interval of 30 Myrs, (b) the distribution of these groups on
the $z,t$ plane, as well as (c) the dependence of parameter $q$ on time $t$.
  }
 \label{f1-XYZq-t}
 \end{center} } \end{figure}
 \begin{figure} {\begin{center}
 \includegraphics[width=0.8\textwidth]{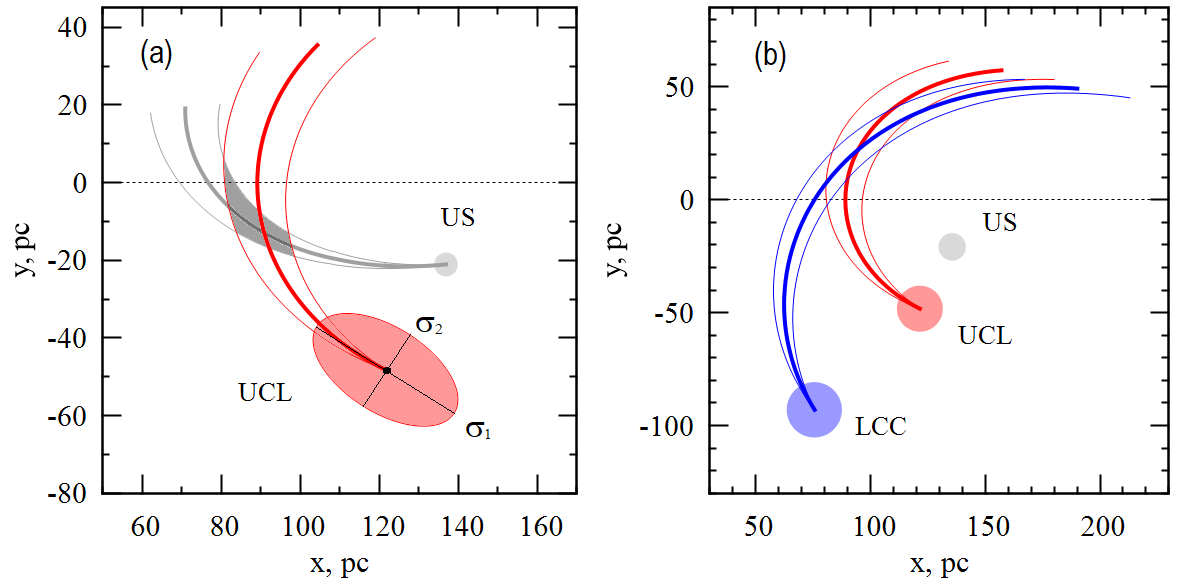}
 \caption{
 (a) Position of US and UCL centers on the $xy$ plane and their trajectories in the past on time interval of 20 million years, the residual velocity ellipsoids are schematically shown, (b) the position of the UCL and LCC centers on the $xy$ plane and their trajectories in the past on time interval of 30 million years, for each trajectory the confidence regions corresponding to the errors of the initial velocities of level 1$\sigma$ are marked.
  }
 \label{f2-UCL-US}
 \end{center} } \end{figure}

\section{RESULTS AND DISCUSSION}
In Table 1, average values of coordinates ${\overline x}_0,{\overline y}_0,{\overline z}_0$ and the average values of the observed velocities relative to the Sun ${\overline U},{\overline V},{\overline W}$ for US, UCL, and LCC centers are given. The number of stars used to calculate the average coordinate values $N_{x,y,z}$ and average velocity values $N_{U,V,W}$ is indicated. Note that for US, there is one peculiarity in calculating coordinates.
Specifically, the table provides coordinates calculated
using 1250 stars. If we calculate them based on 377 stars with known radial velocities, then the ${\overline z}_0$ value turns out to be by 10 pc greater, i.e., ${\overline z}_0=51.4$~pc. The table also gives the initial velocities of the US, UCL, and LCC centers, ${\overline U_0},{\overline V_0},{\overline W_0}$, that were used to construct epicyclic orbits.

In Table 2, average values of coordinates ${\overline x}_0,{\overline y}_0,{\overline z}_0$
and the average values of the observed velocities relative to the Sun ${\overline U},{\overline V},{\overline W}$ for US, UCL, and LCC centers are given. Here, the coordinates and velocities
were obtained by other authors. In [24, 36], the coordinates
and velocities are calculated based on bright
stars from the Hipparcos catalogue. Differences between our results for US velocities and the conclusions of [29] are negligible and do not exceed 0.2 km/s for each component. It is important to note the difference of $\Delta {\overline U}_0\sim2$~km/s for US between our estimates and the results of [24]. This difference already has a noticeable impact on the nature of the US orbit. For example, with the initial data of [24] at [25,] US,
UCL, and LCC orbits were constructed.

In Figs. 1a and 1b, the of US, UCL, and LCC trajectories traced over time interval of 30 million years in the past are given. We can see that there are intersections
of trajectories on the $xy$ plane (Fig. 1a), and in
the vertical direction (Fig. 1b), all three trajectories are
almost parallel. Therefore, it is clear that there were
moments in time at which the mutual distances
between the centers of the groups were minimal. Note
that the trajectories of the stars are calculated taking
into account the elevation of the Sun above the galactic plane. Thus, in our drawings, the $z$ coordinate reflects the position of stars relative to the plane of the Galaxy.

Our US, UCL, and LCC orbits differ from the
orbits constructed in [25]. These differences are
mainly due to two reasons: (a) the difference in the initial
velocities, which affects only the US orbit and (b)
different values of Sun's velocity $(U,V,W)_\odot$. In this
paper, the following parameters (3), where there is a
difference of $\Delta V_\odot\sim$6 km/s from velocity $V_\odot$ from [25]
that was used in [37] were used. As a result, compared
with the US, UCL, and LCC orbits from [25], the centers
of our objects have a noticeable movement along
the $x$ axis and move more slowly along the y axis.

Based on the constructed US, UCL, and LCC trajectories, we calculated parameter using mutual differences of coordinates $\Delta x,\Delta y,\Delta z$ of forms ``UCL minus US'' and ``UCL minus LCC'' for each moment of integration:
 \begin{equation}
 q=\sqrt{\Delta x^2+\Delta y^2+\Delta z^2}.
 \label{qq}
 \end{equation}
Thus, parameter $q$ characterizes the deviation of the
analyzed trajectory from the reference one, where the
UCL trajectory acts as the reference.

Figure 1c shows the dependence of parameter $q$ on
time $t$, where parameter $q$ calculated relative to the
UCL orbit. As follows from a detailed analysis of the
data presented in the graph, the minimal distance
between the UCL and LCC centers was $q_{\rm min}=14.6\pm0.7$~pc
at $t_{\rm min}=-17.7\pm2.4$~million
years. In this case, the minimal distance between the
centers of UCL and US was $q_{\rm min}=21.9\pm1.0$~pc at the
moment of $t_{\rm min}=-6.4\pm1.7$~million years. Parameter
definition errors $q_{\rm min}$ and $t_{\rm min}$ were determined as a
result of statistical modeling using the Monte Carlo
method under the assumption that the individual US,
UCL and LCC orbits are constructed with relative
errors of 10\% and distributed according to the normal
law. We can consider that the kinematic age of the
UCL and LCC is $17.7\pm2.4$~million years and the
kinematic age of US is $6.4\pm1.7$~million years.

It is interesting to note the work [38], where, based on data from the Gaia DR2 catalogue and using the epicyclic approximation to construct stellar trajectories in the Sco-Cen association, estimates of the kinematic ages of its various components were obtained.
These authors identified eight kinematically distinct
components in the association using over 8000 stars.
UCL and LCC were split into two parts each. Estimates
were obtained for two UCL components of $15\pm3$ and $13\pm8$~million years. For the two LCC components, these authors found $7\pm5$ and $9\pm4$~million years. Overall, we have good agreement with the estimates we obtained for these components of the Sco-Cen association.

In Table 3, the parameters of the residual velocity ellipsoid found from two components $V_l$ and $V_b$  using mthree equations (6)--(8) are given. The number of used
stars corresponds to the value of $N_{x,y,z}$ indicated in
Table 1. The last column of the table gives the value of
average spatial dispersion $\sigma_{3D}$, which is calculated as
follows:
 \begin{equation}
 \sigma_{3D}=\sqrt{(\sigma^2_1+\sigma_2^2+\sigma_3^2)/3}.
 \label{sigma3D}
 \end{equation}
In Table 4, the parameters of the residual velocity
ellipsoid found for all three components $V_l$, $V_b$ and $V_r$
using six equations (6)--(11) are given.

In Fig. 2, the positions of the US, UCL, and LCC
centers on the $xy$ plane are given. In Fig. 2a, the US
and UCL trajectories are given in time interval of
20 million years in the past in order to see more clearly
the area of their intersection. Even taking into account
data errors, it is clear that the area of intersection of
US and UCL is very compact. The moment of intersection
of the trajectories is in agreement with the estimates
of the US age. It is important to keep in mind
that the most accurate parameters of the intersection
of the US and UCL trajectories are obtained from the
analysis of the three-dimensional figure (Fig. 1c).

Figure 2a shows a schematic representation of the
residual velocity ellipsoid. It can be noted that the orientation
of the ellipsoid shown in the figure for UCL
corresponds to the $L_1$ values indicated in Tables 3 and
4. Although it should be noted that the direction of the
first axis is determined with large errors.

  \begin{table}[t]
  \caption[]{\small
Parameters of the residual velocity ellipsoid found from only two components $V_l$ and $V_b$ using three equations (\ref{EQsigm-1})--(\ref{EQsigm-3})
 }
  \begin{center}  \label{T-3}    \small
  \begin{tabular}{|r|c|c|c|r|r|r|c|c}\hline
   & $\sigma_1$ & $\sigma_2$ & $\sigma_3$ &
          $L_1;B_1$ & $L_2;B_2$ & $L_3;B_3$ &  $\sigma_{3D}$ \\
    & km/s & km/s & km/s & deg & deg & deg &  km/s \\\hline

 US &$7.07\pm0.49$& $1.93\pm0.41$& $0.40\pm1.08$& 332;~3& $63;16$& $233\pm6;74\pm3$& $4.24\pm0.92$\\
UCL &$7.08\pm1.28$& $2.09\pm0.57$& $0.79\pm1.31$& 297;19& $42;36$& $193\pm9;61\pm2$& $4.24\pm1.01$\\
LCC &$6.73\pm0.62$& $2.31\pm2.71$& $0.50\pm0.90$& 302;~9& $33;~9$& $165\pm15;77\pm5$& $4.12\pm0.56$\\
 \hline
 \end{tabular}\end{center}
 \end{table}
  \begin{table}[t]
  \caption[]{\small
Parameters of the residual velocity ellipsoid found for all three components $V_l$, $V_b$ and $V_r$ using six equations (\ref{EQsigm-1})--(\ref{EQsigm-6})
 }
  \begin{center}  \label{T-4}    \small
  \begin{tabular}{|r|c|c|c|r|r|r|c|c}\hline
   & $\sigma_1$ & $\sigma_2$ & $\sigma_3$ &
          $L_1;B_1$ & $L_2;B_2$ & $L_3;B_3$ &  $\sigma_{3D}$ \\
    & km/s & km/s & km/s & deg & deg & deg & km/s \\\hline

  US &$~9.07\pm0.35$&$1.52\pm0.29$&$0.39\pm1.09$&337;~7&$69;~36$&$234\pm3;53\pm2$& $5.31\pm1.33$\\
 UCL &$11.37\pm0.74$&$4.26\pm0.23$&$0.50\pm2.05$&329;14&$58;-6$&$126\pm4;74\pm2$& $7.02\pm1.64$\\
 LCC &$11.68\pm0.34$&$1.87\pm3.20$&$0.98\pm0.46$&305;~9&$35;~~2$&$141\pm9;78\pm2$& $6.80\pm1.60$\\
 \hline
 \end{tabular}\end{center}
 \end{table}

In Fig. 2b, the UCL and LCC trajectories are given in time interval of 30 million years in the past. To construct the confidence regions, we calculated the average errors of initial velocities ${\overline U}_0,{\overline V}_0,{\overline W}_0$ in
advance depending only on the errors in the measurements
of the observed velocities $V_l,V_b,V_r$. The
values of such errors (level 1$\sigma$) turned out to be the following:
0.45, 1.11, and 0.91 km/s for US, UCL, and LCC, respectively. Here too, the moment of intersection of the UCL and LCC trajectories at $\sim$15 million
years ago is in good agreement with known estimates
of their age obtained not on the basis of kinematics.

The value of the dispersion of internal velocities in
clusters and associations is known to be an indicator of
their dynamic state and depends on the age of the cluster.
The US age, on one hand, and the UCL and LCC
age, on the other, differ by approximately two times.
Found $\sigma_{3D}$ values increase from US to UCL and
LCC, which is evident at least from Table~4.

From Table 4, it can also be seen that the direction of the third ellipsoid axis of residual velocity $L_3$ and $B_3$ is determined with minor errors. The most interesting
thing here is the deviation from the vertical at an angle
of $\sim$12--15$^\circ$, which is usually associated with the orientation
of the third axis of the Gould belt (see, for example, [39]).

\section{CONCLUSIONS}
The kinematics of young stars in the Sco-Cen association
has been studied. For this purpose, stars with
trigonometric parallaxes and proper motions from the
Gaia EDR3 catalogue were used. The selection was
made from the catalog [29], where for some stars, the
radial velocity values collected from literary sources
are given. We focus on studying the kinematics and
statistical properties of the three main components of
the Sco-Cen association: US, UCL, and LCC.

From the analysis of epicyclic orbits of the US,
UCL and LCC centers that are constructed in the
past, the minimal distance between the UCL and LCC
centers has been shown to be $q_{\rm min}=14.6\pm0.7$~pc at
time of $t_{\rm min}=-17.7\pm2.4$~million years while the
minimal distance between the centers of UCL and US
to be $q_{\rm min}=21.9\pm1.0$~pc at time of
$t_{\rm min}=-6.4\pm1.7$~million years. Thus, kinematic estimates of the age
of three components of the Sco-Cen association,
where the age of the UCL and LCC has been equal to
$17.7\pm2.4$~million years and the age of the US has been
$6.4\pm1.7$~million years, have been obtained.

The parameters of the residual velocity ellipsoid
have been estimated. These parameters have been
shown to be found with smaller errors using only two
velocity components $V_l$ and $V_b$. As a result, the following
values of the principal axes of the residual velocity
ellipsoid have been obtained:
 $\sigma_{1,2,3}=(7.07,1.93,0.40)\pm(0.49,0.41,1.08)$ km/s for US,
 $\sigma_{1,2,3}=(7.08,2.09,0.79)\pm(1.28,0.57,1.31)$ km/s for UCL,
 $\sigma_{1,2,3}=(6,73,2.31,0.50)\pm(0.62,2.71,0.90)$ km/s for LCC.

\subsubsection*{ACKNOWLEDGMENTS}
The authors thank the reviewer for useful comments that
contributed to improving the work.

\subsubsection*{FUNDING}
This work was supported by ongoing institutional funding.
No additional grants to carry out or direct this particular
research were obtained.

\subsubsection*{CONFLICT OF INTEREST}
The authors of this work declare that they have no conflicts of interest.

 \subsubsection*{REFERENCES}
 \small

\quad~~1. A. Blaauw, Publ. Kapteyn Astron. Lab. Groningen 52, 1 (1946).

2. A. Blaauw, Ann. Rev. Astron. Astrophys. 2, 213 (1964)

3. P. T. de Zeeuw, R. Hoogerwerf, J. H. J. de Bruijne, A. G. A. Brown, and A. Blaauw, Astron. J. 117, 354 (1999).

4. N. J. Wright and E. E. Mamajek, Mon. Not. R. Astron. Soc. 476, 381 (2018).

5. E. E. Mamajek, M. Meyer, and J. Liebert, Astron. J. 124, 1670 (2002).

6. M. J. Sartori, J. R. D. L\'epine, and W. S. Dias, Astron.
Astrophys. 404, 913 (2003).

7. T. Preibisch and E. Mamajek, in Handbook of Star Forming Regions, Vol. 2: The Southern Sky, Vol. 5 of ASP Monograph Publ., Ed. by Bo Reipurth (Astron. Soc. Pacif., San Francisco, 2008), p. 235.

8. J. H. J. de Bruijne, Mon. Not. R. Astron. Soc. 310, 585 (1999).

9. S. Ratzenb\"ock, J. E. Gro{\ss}schedl, T. M\"oller, J. Alves,
I. Bomze, and S. Meingast, Astron. Astrophys. 677, A59 (2023).

10. S. Ratzenb\"ock, J. E. Gro{\ss}schedl, J. Alves, N. Miret-Roig, et al., Astron. Astrophys. 678, A71 (2023).

11. T. Preibisch and H. Zinnecker, Astron. J. 117, 2381 (1999).

12. E. J. de Geus, Astron. Astrophys. 262, 258 (1992).

13. B. Fuchs, D. Breitschwerdt, M. A. de Avillez, C. Dettbarn,
and C. Flynn, Mon. Not. R. Astron. Soc. 373, 993 (2006).

14. A. Blaauw, Bull. Astron. Inst. Netherland 11, 414 (1952).

15. V. V. Bobylev and A. T. Bajkova, Astron. Lett. 33, 571 (2007).

16. B. Goldman, S. R\"oser, E. Schilbach, A. C. Mo\'or, and
T. Henning 868, 32 (2018).

17. V. V. Bobylev and A. T. Bajkova, Astron. Rep. 64, 326 (2020).

18. V. V. Bobylev and A. T. Bajkova, Astron. Lett. 49, 410 (2023).

19. N. J. Wright and E. E. Mamajek, Mon. Not. R. Astron. Soc. 476, 381 (2018).

20. E. E. Mamajek and C. P. M. Bell, Mon. Not. R. Astron. Soc. 445, 2169 (2014).

21. R. A. Lee, E. Gaidos, J. van Saders, G. A. Feiden, and
J. Gagn\'e, Mon. Not. R. Astron. Soc. 528, 4760 (2024).

22. C. Ducourant, R. Teixeira, P. A. B. Galli, J. F. Le Campion, A. Krone-Martins, B. Zuckerman, G. Chauvin, and I. Song, Astron. Astrophys. 563, A121 (2014).

23. K. L. Luhman, Astron. J. 165, 269 (2023).

24. M. J. Sartori, J. R. D. L\'epine, and W. S. Dias, Astron. Astrophys. 404, 913 (2003).

25. D. Fern\'andez, F. Figueras, and J. Torra, Astron. Astrophys. 480, 735 (2008).

26. ESA, The Hipparcos and Tycho Catalogues. Astrometric and Photometric Star Catalogues derived from the ESA Hipparcos Astrometry Mission, ESA SP-1200 (1997), Vol. 1--17.

27. T. Prusti, J. H. J. de Bruijne, A. G. A. Brown, A. Vallenari,
et al., Astron. Astrophys. 595, A1 (2016).

28. A. G. A. Brown, A. Vallenari, T. Prusti, J. H. J. de Bruijne,
et al., Astron. Astrophys. 649, A1 (2021).

29. K. L. Luhman, Astron. J. 163, 24 (2022).

30. B. Lindblad, Ark. Mat. Astron. Fys. 20A (17), 7 (1927).

31. V. V. Bobylev and A. T. Bajkova, Astron. Lett. 42, 1 (2016).

32. R. Sch\"onrich, J. Binney, and W. Dehnen, Mon. Not. R. Astron. Soc. 403, 1829 (2010).

33. J. Holmberg and C. Flinn, Mon. Not. R. Astron. Soc. 352, 440 (2004).

34. O. I. Krisanova, V. V. Bobylev, and A. T. Bajkova, Astron.
Lett. 46, 370 (2020).

35. K. F. Ogorodnikov, Dynamics of Stellar Systems, Ed. by
A. Beer (Pergamon, Oxford, 1965).

36. S. Madsen, D. Dravins, and L. Lindegren, Astron. Astrophys. 381, 446 (2002).

37. W. Dehnen and J. Binney, Mon. Not. R. Astron. Soc. 298, 387 (1998).

38. M. \u{Z}erjal, M. J. Ireland, T. D. Crundall, M. R. Krumholz,
and A. D. Rains, Mon. Not. R. Astron. Soc. 519, 3992 (2023).

39. V. V. Bobylev, Astrophysics 57, 583 (2014).

 \end{document}